%% file: main.tex
\documentclass[11pt, a4paper, logo, copyright]{deepseek}

\pdftrailerid{redacted}

\makeatletter
\renewcommand\bibentry[1]{\nocite{#1}{\frenchspacing\@nameuse{BR@r@#1\@extra@b@citeb}}}
\makeatother

\input{preamble}

\newcommand{\titlelogo}{\raisebox{-0.22\height}{\includegraphics[height=1.8em]{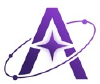}}}

\begin{document}







\begin{titlepage}
\thispagestyle{empty}

\noindent\includegraphics[height=2em]{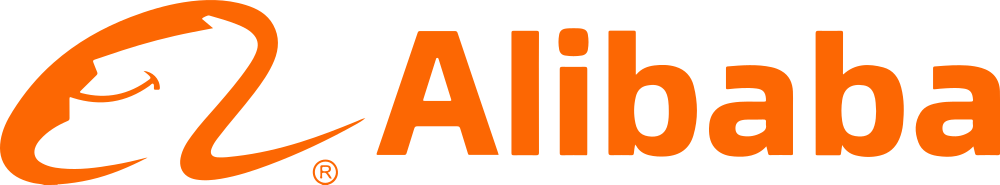}\par
\noindent\rule{\textwidth}{0.2pt}

\vspace{0.1em}

\begin{center}
{\LARGE\bfseries
\titlelogo\hspace{0.45em}
Astar: Learning to Propose Evolution Directions for Self-Evolving Industrial AI Systems
}
\end{center}



\begin{center}
{
\begin{tabular}{@{}l l l l l @{}}
Jinxin Hu$^{1,\dag,*}$ & Hao Deng$^{1,*}$ & Haibo Xing$^{1,*}$ & Lingyu Mu$^{1,*}$ & Muyu Zou$^{1,*}$ \\
Weiqin Yang$^{2}$ & Sirui Chen$^{2}$ & Bohao Wang$^{2}$ & Zhezheng Hao$^{2}$ & Hao Zhang$^{1}$ \\
Zulong Chen$^{1}$ & Shizhun Wang$^{1}$ & Yu Zhang$^{1}$ & Xiaoyi Zeng$^{1}$ & Jiawei Chen$^{2,\dag}$ \\
\end{tabular}
}
\end{center}
\vspace{-2em}
\begin{center}
{\small
$^{1}$ Alibaba Group \quad
$^{2}$ Zhejiang University
}
\end{center}

\vspace{0.25em}


\begin{center}
{\Large\bfseries Abstract}
\end{center}


\begin{adjustwidth}{0.9cm}{0.9cm}
\noindent
\setlength{\parindent}{1em}
\setlength{\parskip}{0pt}
\input{sections/0_abstract}

\end{adjustwidth}


\vfill
\noindent\rule{\textwidth}{0.4pt}

\vspace{0.25em}

\noindent{\footnotesize
$^{*}$ Equal core contribution,\quad
$^{\dag}$ Corresponding author
}



\end{titlepage}

\pagestyle{fancy}
\fancyhf{}
\fancyhead[C]{}
\fancyfoot[C]{\footerfont \thepage}
\renewcommand{\headrulewidth}{1pt}
\renewcommand{\footrulewidth}{1pt}

\setcounter{tocdepth}{2}
\begingroup
\hypersetup{linkcolor=black}
\color{black}
\renewcommand{\baselinestretch}{1.06}\normalsize
\setlength{\parskip}{5.2pt plus 0.8pt}
\tableofcontents
\endgroup
\clearpage

\input{sections/1_introduction}

\begingroup
\setlength{\parskip}{1.4pt plus 0.4pt}%

\input{sections/2_overview}
\input{sections/3_data_v3}
\input{sections/4_posttraining}

\input{sections/5_experiments}

\input{sections/6_application}
\input{sections/7_conclusion}
\endgroup

\clearpage
\bibliographystyle{unsrtnat}
\bibliography{reference}

\clearpage
\appendix
\input{sections/G_data}
\input{sections/B_implementation}

\input{sections/F_system}
\input{sections/D_cases}
\input{sections/A_authors}

\end{document}

%% file: preamble.tex
\usepackage{amsmath}
\usepackage{amsfonts}
\usepackage{mathtools}
\usepackage{bm}
\usepackage{nicefrac}

\usepackage{float}
\usepackage{placeins}
\usepackage{wrapfig}
\usepackage{subcaption}
\usepackage{capt-of}
\usepackage[rightcaption]{sidecap}
\usepackage{array}
\usepackage{multirow}
\usepackage{makecell}
\usepackage[flushleft]{threeparttable}
\usepackage{arydshln}
\usepackage{siunitx}
\usepackage{tabularray}
\UseTblrLibrary{booktabs}

\newcolumntype{C}{>{\centering\arraybackslash}X}
\newcolumntype{L}{>{\raggedright\arraybackslash}X}

\usepackage{algorithm}
\usepackage{algorithmicx}
\usepackage{algpseudocode}
\usepackage{listings}
\setlist[itemize]{
  leftmargin=1.4em,
  labelwidth=0.8em,
  labelsep=0.45em,
  labelindent=0pt,
  topsep=0pt,
  partopsep=0pt,
  itemsep=0.18em plus 0.03em,
  parsep=0pt
}
\setlist[itemize,2]{
  leftmargin=1.55em,
  label={\textbullet},
  topsep=0pt,
  partopsep=0pt,
  itemsep=0.1em plus 0.02em,
  parsep=0pt
}
\setlist[enumerate]{
  leftmargin=1.65em,
  labelwidth=1.0em,
  labelsep=0.45em,
  labelindent=0pt,
  topsep=0pt,
  partopsep=0pt,
  itemsep=0pt,
  parsep=0pt
}

\usepackage{gdm-colors}
\usepackage[most]{tcolorbox}
\usepackage{soul}
\usepackage{xcolor}

\definecolor{thinkcolor}{RGB}{227,196,144}
\definecolor{observecolor}{RGB}{153,201,227}
\definecolor{explorecolor}{RGB}{178,217,200}
\definecolor{taomindlink}{RGB}{158,74,18}
\definecolor{taomindcite}{RGB}{24,62,112}
\definecolor{taomindurl}{RGB}{48,78,88}

\usepackage{tikz}
\usetikzlibrary{shapes.geometric, arrows.meta, positioning}

\usepackage{multicol}
\usepackage{calc}
\usepackage[bottom]{footmisc}

\usepackage{fontawesome}

\hypersetup{
    colorlinks=true,
    linkcolor=taomindlink,
    citecolor=taomindcite,
    urlcolor=taomindurl,
    filecolor=taomindlink,
    pdfborder={0 0 0},
    breaklinks=true
}
\usepackage[nameinlink]{cleveref}
\crefname{table}{Table}{Tables}
\Crefname{table}{Table}{Tables}
\crefname{figure}{Figure}{Figures}
\Crefname{figure}{Figure}{Figures}
\crefname{section}{Section}{Sections}
\Crefname{section}{Section}{Sections}
\crefname{equation}{Equation}{Equations}
\Crefname{equation}{Equation}{Equations}

\usepackage[numbers, sort&compress, square]{natbib}

\usepackage[normalem]{ulem}
\usepackage{dashrule}
\usepackage{blindtext}
\usepackage{tablefootnote}
\usepackage{xspace}

\newcommand{\modelicon}[1]{\raisebox{-0.15em}{\includegraphics[height=0.9em]{icon/#1.pdf}}}

\newcounter{caseexample}[section]

\newcounter{promptexample}[section]

%% file: sections/0_abstract.tex
Modern AI systems advance through continuous iteration: a loop of proposing evolution directions, implementing code, training, and evaluation. While the latter three stages are increasingly automated, the starting point --- proposing effective evolution directions --- remains a critical bottleneck that still relies heavily on senior experts. In this work, we explore whether AI can take over this role.  We find that general-purpose LLMs, even the advanced GPT-5.5, offer only generic and misaligned suggestions:  the required expertise is accumulated through experience rather than explicitly codified, and thus hard to inject directly. 

To this end, we propose \textbf{Astar}, a training-based approach that learns a specialized evolution-guiding model from the abundant iteration histories of industrial systems. Realizing this idea, however, raises four challenges: \emph{sparse supervision}, \emph{noisy data}, \emph{a vast direction space}, and \emph{prohibitively expensive verification}. We address them along two fronts. On the data side, we design a pipeline that turns noisy historical commits into a large, clean evolutionary corpus via pairwise sample expansion and noise filtering. On the model side, we train the model through mid-training, SFT, and RL, guiding evolution direction generation with hierarchical hints and using the reward model in RL as a fast surrogate evaluator.

Astar has been deployed in Alibaba's Lazada advertising system for evolution direction proposal. Astar-8B achieves a single-proposal success rate of 0.6786 in real-execution evaluation, far exceeding human experts (0.3229) and the strongest general-purpose LLM (0.3071).  More importantly, Astar closes the loop and enables fully automatic iteration: it guided 20 consecutive iterations over two weeks, improving offline Hitrate@200 by 23.6\%, while an online A/B test yielded relative lifts of 4.86\% in GMV and 1.82\% in advertising revenue.

%% file: sections/1_introduction.tex
\section{Introduction}
\label{sec:intro}

\begin{figure}[t]
  \centering
  \includegraphics[width=1.0\linewidth]{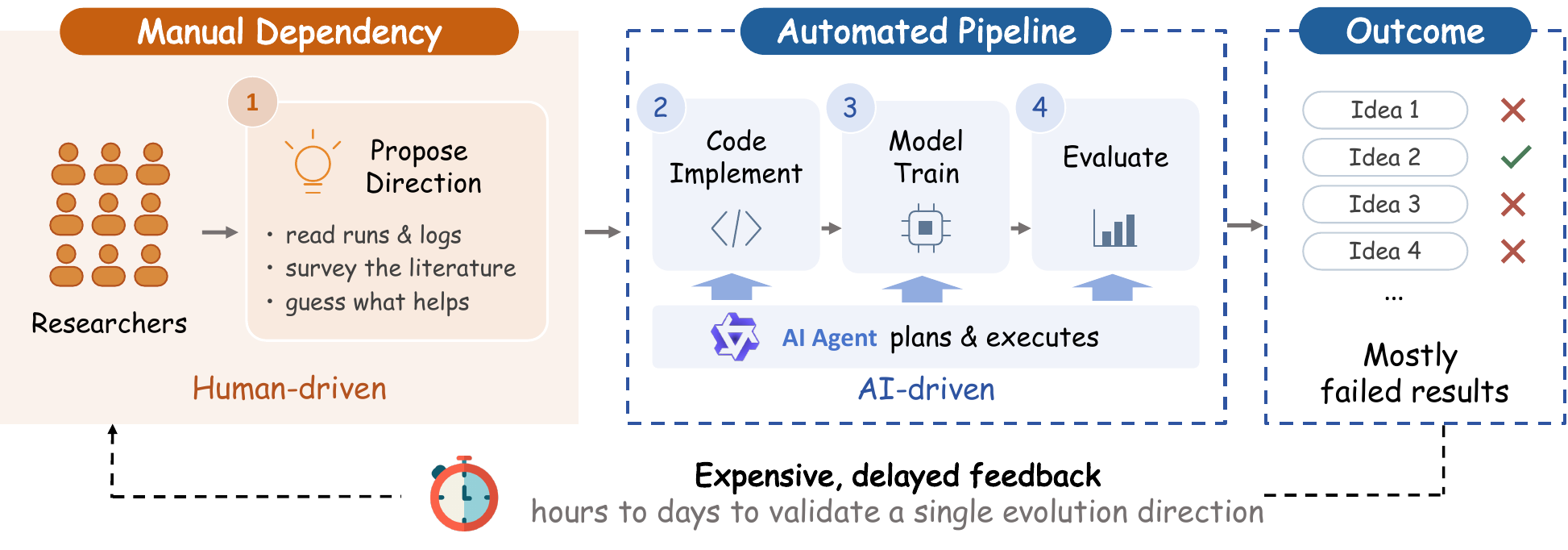}
  \caption{The iteration cycle of modern AI systems.}
  \label{fig1}
  \vspace{-0.5cm}
\end{figure}

Modern AI models have been widely applied across diverse domains, including e-commerce~\citep{zhai2024actions, li2024ecomgpt, deng2025onerec}, healthcare~\citep{singhal2025toward, tu2025towards, ding2025multimodal}, and finance~\citep{cao2024man, kelly2024virtue, duarte2024machine}.
In real-world industrial practice, their success stems not from a one-time architecture design, but increasingly from continuous iteration --- a loop of four stages: \emph{proposal of evolution directions}, \emph{code implementation}, \emph{model training}, and \emph{downstream evaluation} (Figure~\ref{fig1}).  Within this loop, the evaluation results determine whether each proposal is retained or discarded, thereby driving the continual improvement of AI models.

With the development of LLMs and agent techniques, the latter three stages --- code implementation, model training, and evaluation --- are being rapidly accelerated and automated in industry, with ever less human involvement~\citep{yang2024swe, chan2025mle, cui2026effects}. In contrast, the proposal of evolution directions, the loop's starting point, remains hard to automate and has become the critical bottleneck of automated AI evolution. This stage is difficult for two reasons. First, it is intellectually demanding: formulating a promising direction requires a deep understanding of the business, data, and model characteristics, careful scrutiny of model configurations, training dynamics, and performance, and substantial reasoning about how the model could be improved. Second, it is high-stakes: validating even a single proposal entails a full implementation, training, and evaluation cycle, which often takes several days in typical industrial applications.
Consequently, a poorly chosen direction incurs an enormous waste of time and computational resources. For these reasons, the stage still depends on senior and experienced scientists, and its outcome hinges heavily on their expertise. Given this bottleneck, a question naturally arises: \emph{can AI automatically propose effective evolution directions?} Once answered, the evolution loop could be fully closed and automated, significantly accelerating large-scale industrial AI evolution and fostering the deployment of AI across a wide range of scenarios.

To this end, a natural idea is to directly employ LLMs to propose evolution directions. However, general-purpose LLMs possess limited knowledge of the specific scenario and lack any practical experience with it.
Empirically, we find that their suggestions are consequently generic --- plausible-sounding yet misaligned with the system's actual needs~\citep{si2026ideation,chen2026measuring}.
Although one might inject domain knowledge or expertise through agent techniques such as prompting or skills, the core difficulty is that such knowledge is hard to distill and organize into structured linguistic form --- it is accumulated through experience rather than explicitly codified. 

To fully unlock the ability of LLMs to generate effective evolution directions, we instead pursue a training-based approach. Although formalized knowledge is absent, industrial AI systems routinely log their iteration history as a sequence of commits, which are often readily available and abundant.
Each commit records an evolution step made by engineers: starting from the current model, it captures the proposed change and the resulting shift in performance.
These commits reveal both which directions have been tried and which ones actually worked. By learning from such history, the model can capture the patterns that distinguish effective directions from ineffective ones, thereby acquiring the ability to propose promising directions.

While appealing, this idea is challenging to realize. Unlike other machine learning tasks, training such an evolution-guiding model on iteration history poses four key challenges, spanning both corpus construction and model training:

\begin{itemize}[topsep=3pt]
\item \textbf{(C1) Historical records carry limited direct supervision signals.} Although a system may continuously accumulate extensive version and experiment records, their absolute number is often modest, typically on the order of hundreds. If we consider only the evolution directions actually proposed by engineers --- which occur solely between adjacent commits --- the resulting supervision samples are relatively sparse.

\item \textbf{(C2) Informative samples are buried in noise.} Among the recorded updates, only a fraction involve performance-relevant changes, while many others consist entirely of irrelevant edits such as configuration changes, logging, and refactoring. Training on such raw records leads the model to learn spurious patterns and may even hamper the training process.

\item \textbf{(C3): Vast evolution direction space.} An evolution attempt can target broad dimensions such as model architecture, loss functions, optimizers, and data, etc. Each dimension further branches into fine-grained modifications to specific modules and parameters. Identifying an effective direction without guidance is extremely difficult within such a massive space.

\item \textbf{(C4) Verifying an evolution direction is prohibitively expensive and delayed.} Assessing the effectiveness of a direction requires a full implementation-training-evaluation cycle, costing hours or even days of compute per candidate. This poses significant difficulties for direction exploration and model optimization, which both rely on timely verification feedback.

\end{itemize}

To address these challenges, we build Astar, a framework that trains a specialized model to propose evolution directions. We first construct a evolutionary corpus from raw historical records, tackling the above challenges as follows: 

\begin{itemize}[topsep=3pt]


\item \textbf{To address C1, we expand the samples in a pairwise manner.} Instead of using only adjacent commits, we pair arbitrary two versions in the iteration history to create a quadratic number of training samples: we label which version is better by comparing their full loss curves and summarize the differences between their revisions.

\item \textbf{To address C2, we introduce a noise-filtering strategy.} We first apply rule-based filtering to remove performance-irrelevant samples via reachability analysis, abstract syntax tree (AST) normalization, and configuration whitelisting. 
We then employ LLMs for finer-grained semantic analysis, inferring the explicit evolution intent from code differences and discarding noisy samples whose intent is ambiguous.
\end{itemize}

Building on this corpus, we construct a specialized LLM-based model through the standard stages of mid-training, supervised fine-tuning (SFT), and reinforcement learning (RL). Mid-training enables the model to generate structured evolution directions; SFT further encourages it to produce effective directions that lead to model improvements; and RL is finally employed to enhance the model's generalization and exploration toward novel directions. In particular, to tackle challenges C3 and C4, we develop the following strategies: 

\begin{itemize}[topsep=3pt]
\item \textbf{To address C3, we leverage hierarchical hints to facilitate the generation of evolution directions.} Instead of generating directions directly within this vast space, we first instruct the model to produce hierarchical hints following a three-level taxonomy. This design guides the model to progressively narrow the search space: generating coarse-grained directions first, then localizing fine-grained modification actions, and finally producing a concrete modification plan within the identified scope.

\item \textbf{To address C4, we train a reward model as a surrogate evaluator.}
We explicitly train a reward model from historical records to predict whether a candidate change will improve model performance, turning a verification that would take days into a prediction obtained in seconds. Acting as a proxy, this model both supplies the reward signal for reinforcement learning and ranks candidate directions to select the most promising ones.

\end{itemize}

\begin{figure}[t]
  \centering
  \includegraphics[width=1.0\linewidth]{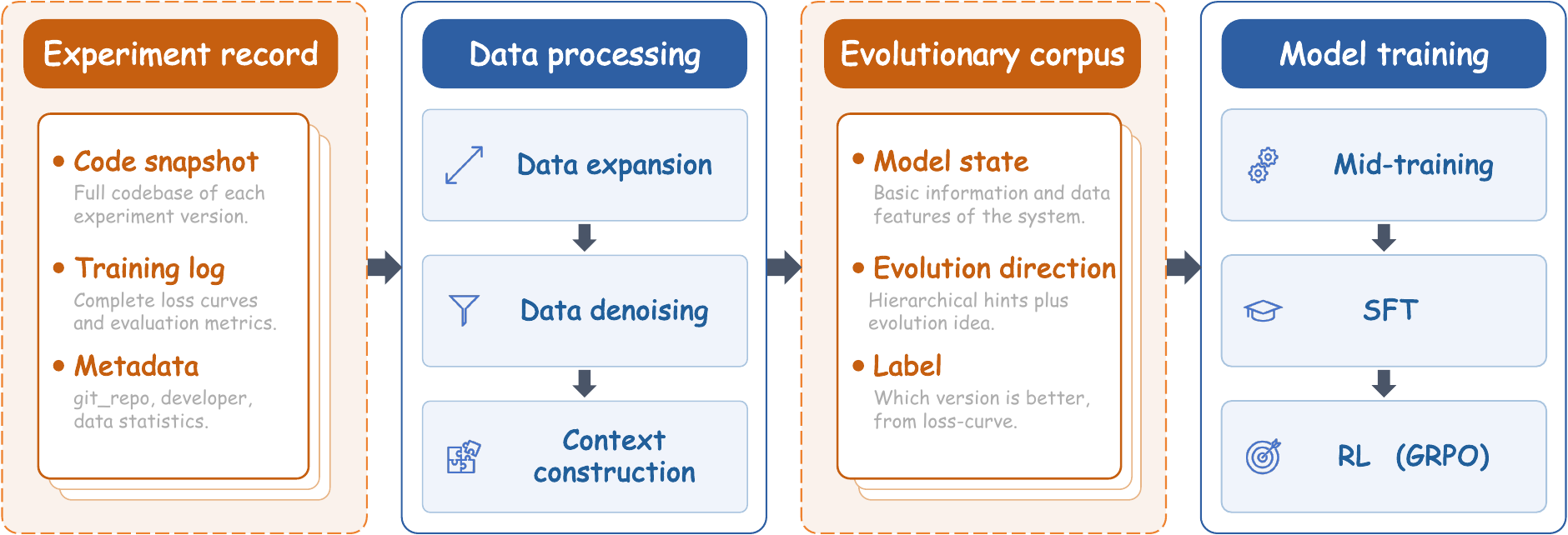}
  \caption{Overall Framework of Astar.}
  \label{fig_process}
  \vspace{-0.5cm}
\end{figure}

To thoroughly validate Astar, we train Qwen3-based models at 0.6B, 4B, and 8B scales and deploy the resulting system on the online recall model of Alibaba’s Lazada advertising platform. In real-execution evaluation, Astar-8B achieves a single-proposal success rate of 0.6786, substantially exceeding 0.3229 for human experts and 0.3071 for the strongest general-purpose LLM. Its reward model reaches an AUC of 0.8487 in predicting whether an evolution direction will improve performance, compared with 0.6142 for human experts and 0.5997 for the best general-purpose LLM. More importantly, Astar guided 20 consecutive iterations over two weeks, improving offline Hitrate@200 by 23.6\%.
An online A/B test further yielded relative lifts of 4.86\% in GMV, 1.82\% in advertising revenue, 0.84\% in clicks, and 1.79\% in orders.

\begin{figure}[t]
  \centering
  \includegraphics[width=1.0\linewidth]{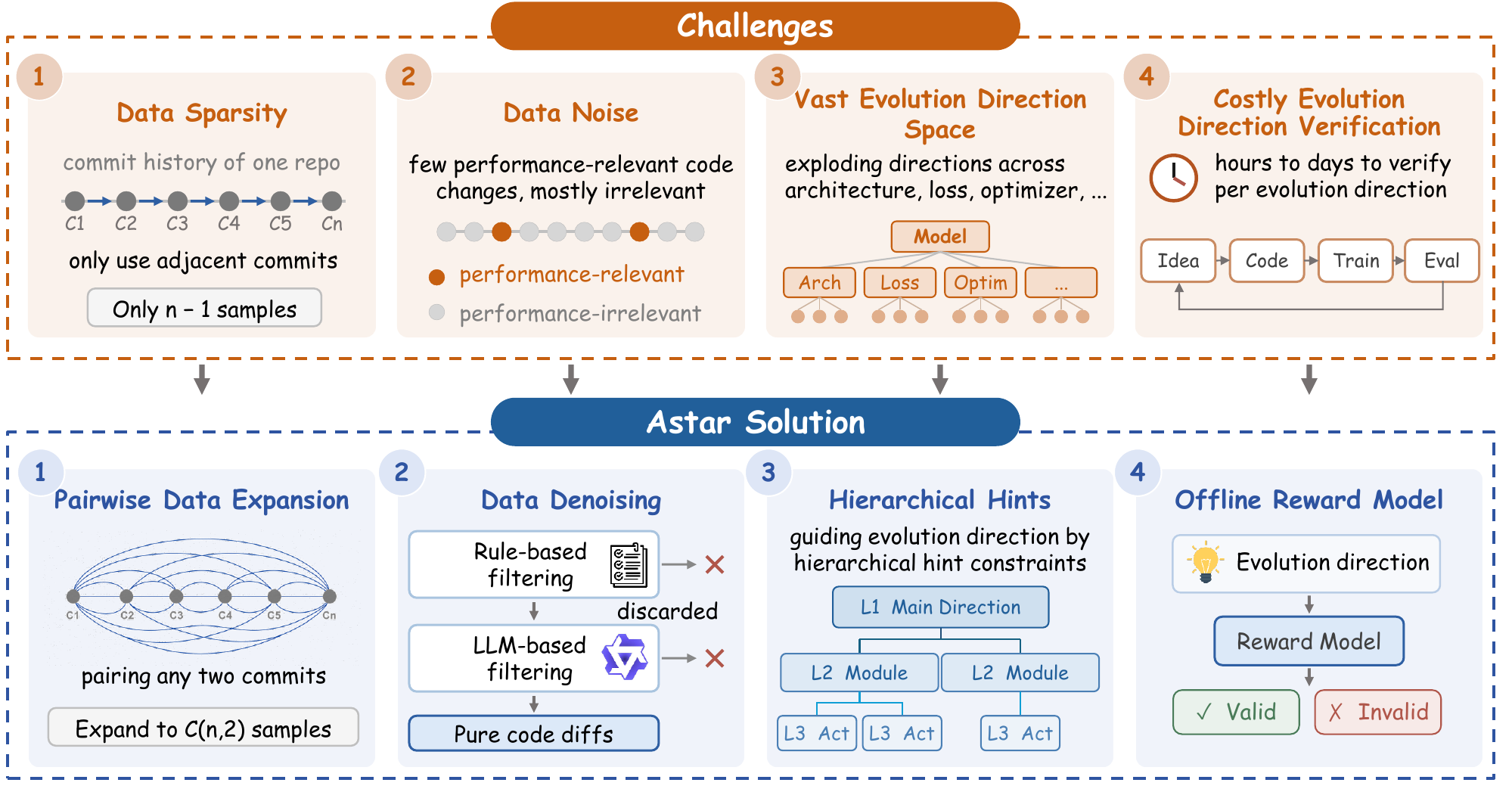}
  \caption{Four challenges in learning evolution directions from iteration history (top) and Astar's four corresponding designs (bottom).}
  \label{fig2}
  \vspace{-0.5cm}
\end{figure}

%% file: sections/3_data_v3.tex
\section{Evolutionary Corpus Construction}
\label{sec:data}

\begin{figure}[t]
    \centering
    \includegraphics[width=1.0\textwidth]{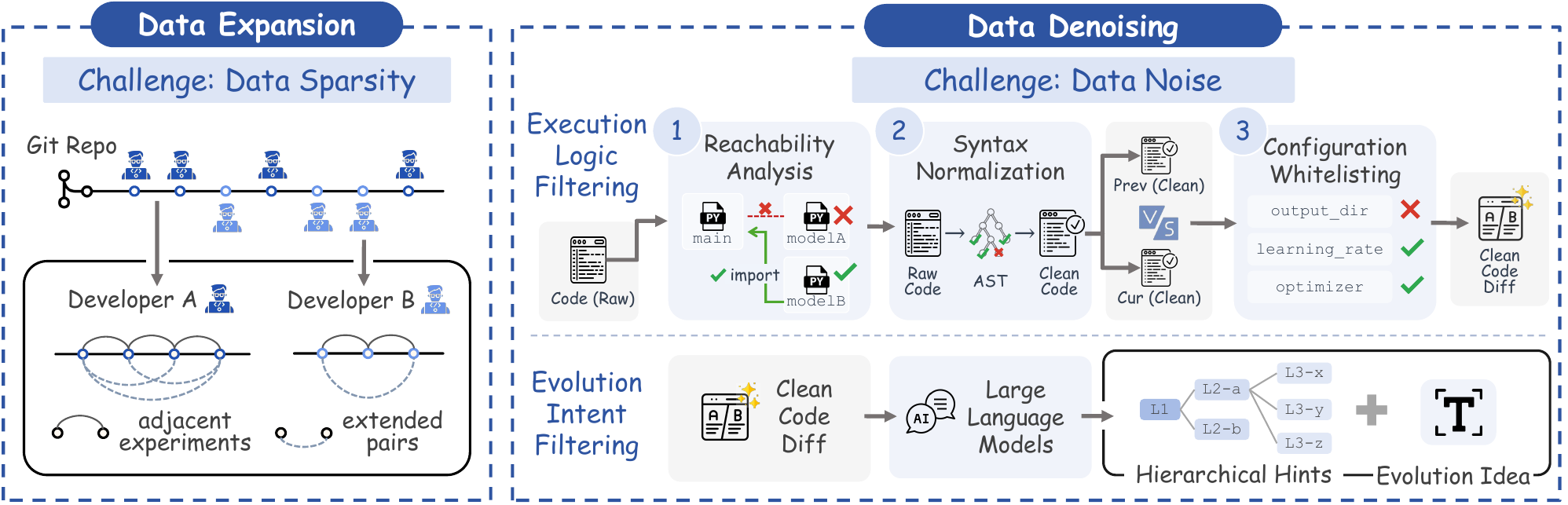}
    \caption{The evolutionary corpus construction pipeline.}
    \label{fig:data_pipeline}
\end{figure}

To enable {Astar} to learn effective model evolution strategies from historical developer trajectories, we construct a specialized evolutionary corpus based on git commits and experimental logs. However, directly utilizing these raw records presents two critical challenges: (1) \textit{Data sparsity} (Challenge C1), where meaningful updates between strictly adjacent experiments are exceedingly rare; and (2) \textit{Data noise} (Challenge C2), where raw code modifications frequently entangle performance-relevant changes with irrelevant engineering edits lacking explicit evolutionary intent. In this section, we introduce a comprehensive data processing pipeline, shown in Figure~\ref{fig:data_pipeline}, to transform raw historical records into a high-quality, structured corpus of state-direction evolutionary pairs.

\subsection{Data Expansion}
\label{sec:data_expansion}
A primary challenge in leveraging historical logs is \textit{data sparsity}. For a given developer within a specific repository (\texttt{git\_repo}), their historical experiments form a chronological sequence $\{E_1, E_2, \dots, E_n\}$. Relying exclusively on strictly adjacent experiments (i.e., $E_t \rightarrow E_{t+1}$) yields insufficient training data and often captures incomplete micro-edits rather than meaningful evolutionary steps (which may span multiple experiments).

To overcome this, we propose a pairwise data expansion strategy. Instead of only extracting consecutive pairs, we generate a quadratic number of combinations by pairing any two experiments $(E_{\mathrm{prev}}, E_{\mathrm{cur}})$. Crucially, for each pair, we compare their full loss curves to objectively label which version achieves better performance (the specific comparison method is detailed in Section~\ref{sec:sft}), determining the direction of improvement. This forms our pariwise training samples. This combinatorial approach transforms a sparse experimental sequence into a dense dataset. Beyond amplifying the sample size, it enables Astar to learn evolutionary trajectories across varying temporal distances, capturing both short-term tweaks and long-term performance leaps.

\subsection{Data Denoising}
\label{sec:data_denoising}
Beyond data sparsity, historical records suffer from severe \textit{noise}. To capture meaningful model evolution, we extract and process evolution-driven code differences rather than the full codebases. This eliminates unchanged boilerplate and forces Astar to focus on actual performance-relevant updates. However, raw deltas remain noisy: they frequently mix core algorithmic changes with irrelevant edits (e.g., logging, refactoring), and many experiments entirely lack explicit evolutionary intent. Thus, we implement a two-stage denoising pipeline comprising rule-based Execution Logic Filtering and LLM-based Evolution Intent Filtering.

\subsubsection{Execution Logic Filtering}
\label{sec:execution_logic_filtering}
This stage aims to isolate modifications that genuinely alter the algorithmic behavior. We achieve this through a sequence of codebase-level cleaning followed by delta-level filtering.

\paragraph{Reachability Analysis.} 
We begin by processing the codebases in the previous experiment $E_{\mathrm{prev}}$ and the current experiment $E_{\mathrm{cur}}$ independently. To ensure the extracted modifications are actually invoked during the model's lifecycle, we construct a call-graph for each codebase. Any code residing in uncalled functions or unreachable branches is identified as dead code and discarded, establishing a strictly execution-relevant foundation.

\paragraph{Syntax Normalization.} 
Continuing at the codebase level, we parse the filtered code into Abstract Syntax Trees (ASTs) to eliminate syntax-level noise that does not affect execution. This step automatically removes trivial components, such as debugging statements (e.g., \texttt{print}, \texttt{assert}) and comments, while normalizing non-functional variations like formatting adjustments and \texttt{import} sequence reordering.

\paragraph{Configuration Whitelisting.} 
After purifying the individual codebases, we employ standard differencing algorithms to compute the pairwise code differences between them. To distinguish between environmental setups and performance-relevant hyperparameters within these extracted changes, we filter the modifications against a predefined parameter whitelist. Changes to engineering variables (e.g., \texttt{output\_dir}, \texttt{gpu\_count}, \texttt{job\_name}) are discarded, while core algorithmic parameters (e.g., \texttt{learning\_rate}, \texttt{optimizer}, \texttt{model\_name}) are preserved, yielding the final purified code delta.

\subsubsection{Evolution Intent Filtering}
\label{sec:evolution_intent_filtering}
Even after obtaining the purified code modifications, certain experimental pairs may still lack a purposeful model evolution intent. To address this, we leverage LLMs to perform fine-grained semantic analysis, mapping the code changes into explicit evolutionary semantics. Modifications that fail to map to valid evolution intents are regarded as ambiguous noise and discarded.

\paragraph{Hierarchical Hints.} 
To tackle the vast space of evolution directions, we use the LLM to classify the purified modifications into a predefined three-level taxonomy: (Level-1) \textit{Coarse-Grained Direction} $\rightarrow$ (Level-2) \textit{Fine-Grained Modification Action} $\rightarrow$ (Level-3) \textit{Concrete Modification Plan}. These \textit{Hierarchical Hints} effectively filter out irrelevant changes while providing Astar with dense, multi-granular supervision signals during training, guiding it to progressively narrow the search space during inference.

\paragraph{Evolution Idea.} 
Complementing the structured taxonomy, we task the LLM with generating an \textit{Evolution Idea}, a concise textual explanation detailing the exact code changes. It captures nuanced algorithmic tweaks that the predefined taxonomy might miss, enriching the dataset with a fine-grained semantic description of the evolution intent.

\subsection{Context Construction}
\label{sec:context_construction}
To predict the optimal evolution direction, Astar must understand the system's state prior to modification. For each sample $(E_{\mathrm{prev}}, E_{\mathrm{cur}})$, we construct a comprehensive system context comprising \textit{Data Features}, which provide structured statistics of the experimental dataset (e.g., business domain, sample volume, feature dimensions) as a quantitative baseline; and a \textit{Basic Information}, a natural language overview of the codebase in the previous experiment $E_{\mathrm{prev}}$, which summarizes the existing model architecture, core modules, and training strategies.

\begin{figure}[t]
    \centering
    \includegraphics[width=1.0\textwidth]{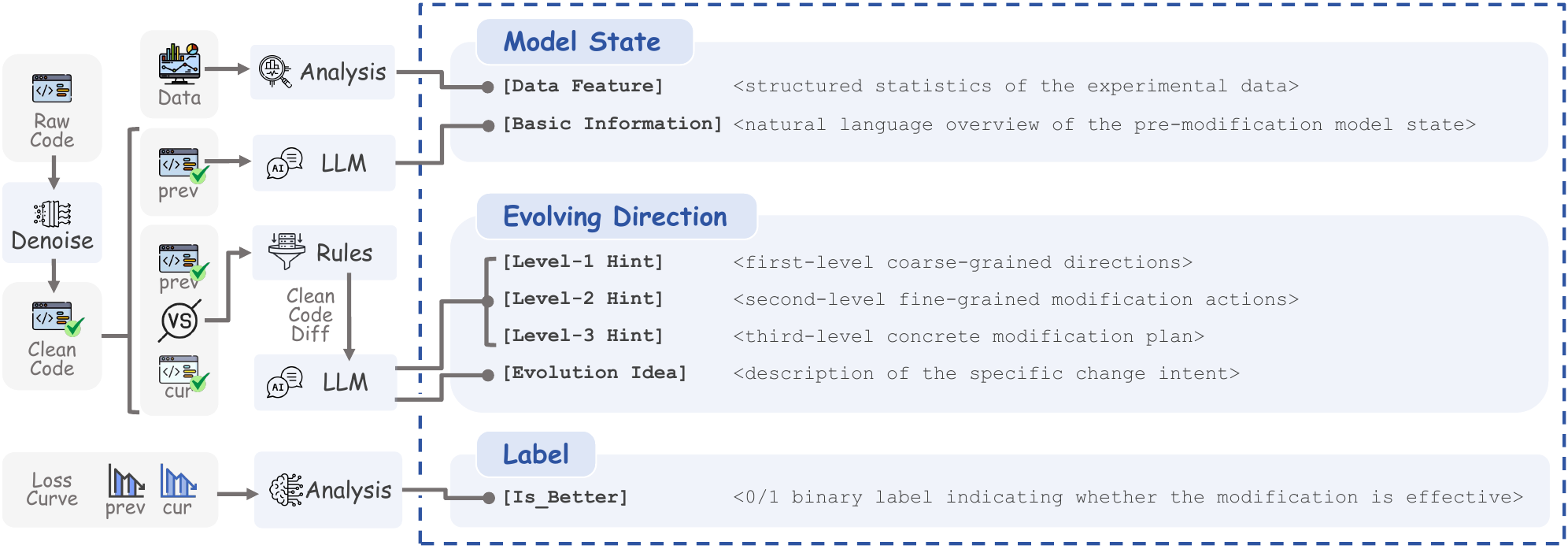}
    \caption{The format of the data.}
    \label{fig:data_format}
\end{figure}

By pairing this state context with the denoised evolutionary signals, we formulate the final training corpus. Each instance structurally maps the initial system state to the intended evolutionary action in the format shown in Figure~\ref{fig:data_format}.

%% file: sections/4_posttraining.tex
\section{Model Training}
\label{sec:post-training}


Training is critical for enabling the model to learn evolution strategies from the history of AI systems. Rather than simply injecting this history into the model's context window --- which risks prompting the model to merely memorize past attempts~\citep{carlini2021extracting} --- training compels the model to internalize intrinsic evolutionary experiences. This allows the model to generalize these experiences to novel scenarios, thereby facilitating the sustainable evolution of AI systems.

\paragraph{Challenges in training.}
However, the training process of Astar faces significant challenges that directly impede the model's ability to learn effective evolution strategies:

\begin{itemize}[topsep=3pt]

    \item \emph{(Challenge C3) Vast evolution direction space.} An evolution attempt can target broad dimensions such as model architecture, loss functions, optimizers, and data, etc. Each dimension further branches into fine-grained modifications to specific modules and parameters, such as selecting the optimizer type, tuning the learning rate, or designing early stopping strategies for optimizers. This makes evolution directions endless, and navigating such a massive space creates substantial uncertainty when the model learns and decides on evolution directions.
    
    \item \emph{(Challenge C4) Expensive verification cost for evolution exploration.} Validating an optimization attempt typically demands substantial computational resources and time~\citep{fetterman2023tune}. Each evolution trial may involve training and validation processes spanning hours or even days, making evolution exploration prohibitively expensive under resource constraints.

\end{itemize}

\paragraph{Our approach.}
To facilitate the model's learning of effective evolution strategies, we propose a three-stage training pipeline, comprising \emph{mid-training}, \emph{supervised fine-tuning} (SFT), and \emph{reinforcement learning} (RL), with specific strategies designed to tackle the above challenges:

\begin{itemize}[topsep=3pt]

\item For \emph{Challenge C3}, we leverage the hierarchical hints extracted during data preprocessing (cf. Figure~\ref{fig:data_format}) to constrain the evolution space. During the mid-training and supervised fine-tuning (SFT)~\citep{Wei2021FinetunedLM,ouyang2022instructgpt} stages, we guide the model to generate structured evolution direction outputs with these hierarchical hints. Consequently, when generating specific evolution directions, the model first produces high-level evolution strategies, progressively narrowing the evolution space and significantly reducing uncertainty in direction selection.

\item For \emph{Challenge C4}, we train a reward model~\citep{Christiano2017DeepRL,Stiennon2020LearningTS} to evaluate the potential of an evolution direction. Using verifiable metrics from historical evolution trajectories (e.g., loss), the reward model learns to predict whether an evolution direction will yield performance improvements (e.g., loss reduction). During reinforcement learning~\citep{Christiano2017DeepRL,Bai2022TrainingAH,rafailov2023dpo}, the reward model provides near-instantaneous feedback for evolution exploration, enabling more efficient search; during inference, it performs rapid ranking of generated evolution direction candidates, substantially lowering verification costs.

\end{itemize}

\subsection{Mid-Training}
\label{sec:mid-training}

General-purpose language models, pretrained on broad corpora, lack exposure to the domain experience of AI system evolution. To address this gap, we introduce a mid-training stage~\citep{liu2025midtraining,faroz2025domain} that aligns the model with the structured representation of evolution directions. Following standard practice in domain adaptation, mid-training performs next-token prediction over the complete evolutionary history dataset (cf. Figure~\ref{fig:data_format}), totaling 112B tokens. Through this process, the model learns to generate evolution directions annotated with hierarchical hints, thereby achieving alignment with the required output format.

\paragraph{Benefits of hierarchical hints.}
The introduction of hierarchical hints offers two key advantages. During training, the model implicitly clusters evolution directions of the same category, reducing overfitting to specific optimization details and improving generalization capability. During inference, the model first generates high-level optimization strategies before elaborating specific modifications, progressively constraining the evolution space and substantially reducing generation uncertainty. This hierarchical decomposition transforms the original massive  search space into a structured, manageable decision process.


\subsection{Supervised Fine-Tuning (SFT)}
\label{sec:sft}

While mid-training enables the model to generate structured evolution directions, the complete evolutionary history contains both successful improvements and failed attempts. To align the model's output distribution toward high-quality evolution directions that yield performance gains, we introduce a supervised fine-tuning stage~\citep{chu2025sft} following mid-training.

\paragraph{Filtering high-quality evolution directions.}
To distinguish high-quality from low-quality evolution attempts, we leverage the loss metric from historical evolution trajectories. Specifically, we define a combined loss metric:
\begin{equation}
s = (1-\alpha)\cdot \ell_T + \alpha \cdot \bar{\ell}_k + \gamma\cdot\hat{\sigma}_k,
\label{eq:conv-score}
\end{equation}
where $\ell_T$ represents the final-step loss, $\bar{\ell}_k$ denotes the mean loss over the final $k$ steps, and $\hat{\sigma}_k$ captures their standard deviation. This formulation comprehensively measures both the convergence outcome and the stability of the training process: the mean term filters transient noise, the final-step term captures the ultimate convergence state, and the stability term penalizes oscillatory behavior.

For each experiment pair $(E_{\mathrm{prev}}, E_{\mathrm{cur}})$ in the training data, we compute their respective loss metrics $s_{\mathrm{prev}}$ and $s_{\mathrm{cur}}$ and evaluate the difference. The label is determined by the loss change:
\begin{equation}
z = \mathbb{1}[s_{\mathrm{cur}} < s_{\mathrm{prev}}],
\label{eq:label}
\end{equation}
where a positive sample ($z=1$) indicates that the evolution attempt yields performance improvement (i.e., $s_{\mathrm{cur}} < s_{\mathrm{prev}}$), and a negative sample ($z=0$) represents a failed attempt. After this filtering process, we retain 81.1M tokens of positive samples and perform standard supervised fine-tuning exclusively on these high-quality evolution directions. Through the SFT stage, the model's output distribution shifts toward evolution directions that demonstrably improve performance, further enhancing the effectiveness of the model's evolution direction generation.

\begin{table}[t]
\caption{Reward-model evaluation against human experts and general-purpose LLMs. The general-purpose LLMs serves as a reward model by scoring and ranking evolution direction candidates. AUC captures the ability to rank positive evolution directions higher than negative ones, while Accuracy is measured at a score threshold of 0.5.}
\label{tab:rm_eval}
\centering
\small
\renewcommand{\arraystretch}{1.05}
\begin{tabular*}{\linewidth}{@{\extracolsep{\fill}} l cc}
\toprule
\textbf{Model} & \textbf{AUC} & \textbf{Accuracy} \\
\midrule
\modelicon{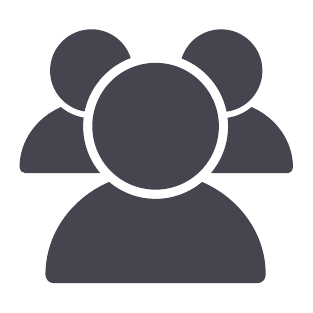} Human & 0.6142 & 0.6039 \\
\midrule
\modelicon{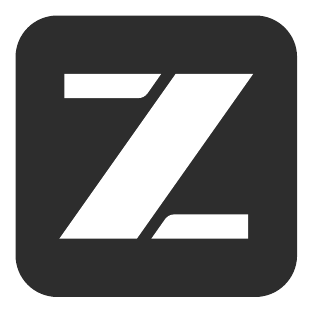} GLM-5.2 & 0.5468 & 0.4844 \\
\modelicon{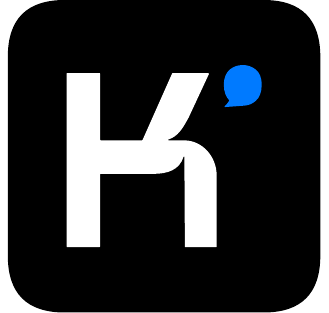} Kimi-K2.6 & 0.5997 & 0.5469 \\
\modelicon{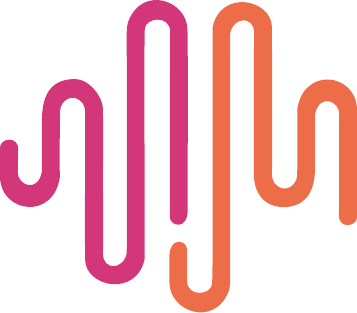} MiniMax-M2.7 & 0.4647 & 0.4141 \\
\modelicon{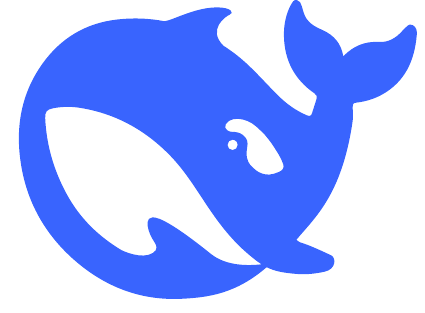} DeepSeek V4 Pro & 0.5982 & 0.5781 \\
\modelicon{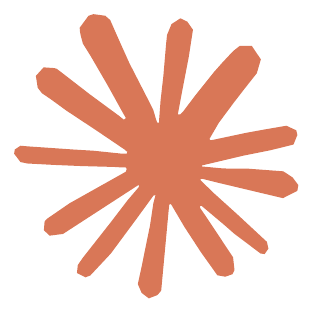} Claude-4.8-Opus & 0.5328 & 0.5469 \\
\modelicon{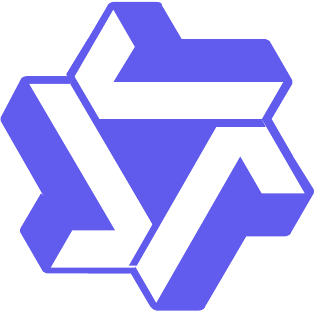} QWen-3.7 Max & 0.5373 & 0.5469 \\
\modelicon{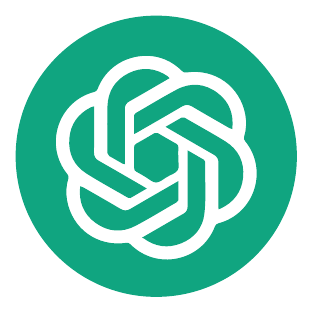} GPT-5.5 & 0.5477 & 0.5078 \\
\midrule
\textbf{Astar-Reward-0.6B} & 0.8094 & 0.7734 \\
\textbf{Astar-Reward-4B} & 0.8253 & 0.7812 \\
\textbf{Astar-Reward-8B} & \textbf{0.8487} & \textbf{0.7890} \\
\bottomrule
\end{tabular*}
\end{table}

\subsection{Reinforcement Learning (RL)}
\label{sec:rl}

While the SFT stage equips the model with the capability to generate high-quality evolution directions, it remains fundamentally constrained by fitting historical successful trajectories. It lacks the ability to explore novel evolution directions unseen in the training data --- a capability essential for the sustained evolution of AI systems. To enhance the model's exploratory capacity, we introduce a reinforcement learning stage. However, verifying the performance impact of an evolution attempt incurs substantial cost, typically requiring complete training and deployment validation cycles. This expense makes it infeasible to provide verifiable feedback for the model's evolution exploration, thereby limiting improvements to the model's exploratory capability.

\paragraph{Reward model for evolution evaluation.}
To overcome this challenge, we first train a reward model~\citep{Christiano2017DeepRL,Stiennon2020LearningTS,lightman2023verify} using the positive and negative samples identified during the SFT stage to evaluate the potential value of evolution directions. Specifically, we replace the model's output layer with a pooling layer followed by an MLP prediction head. For a given evolution direction, the reward model predicts a binary classification probability indicating whether it will yield performance improvement (e.g., loss reduction).

Remarkably, the trained reward model substantially surpasses both human experts and general-purpose LLMs in judging the potential of evolution directions. As shown in Table~\ref{tab:rm_eval}, while human experts achieve an AUC of only 0.6142 and the best general-purpose LLM reaches 0.5997, our reward model attains an AUC of 0.8094 even at 0.6B parameters, rising to 0.8487 at 8B parameters. This demonstrates that the reward model has learned domain-specific patterns from evolutionary history that are inaccessible to general knowledge alone, thereby providing a reliable feedback signal for the model's evolution exploration.

\paragraph{GRPO for evolution exploration.}
With the trained reward model serving as the reward function, we apply the standard GRPO algorithm~\citep{grpo} for reinforcement learning training over 8.06M tokens. Given an AI system state, the model generates a batch of evolution direction candidates (e.g., 16), which the reward model evaluates to compute a reward value for each candidate. The model then performs policy updates based on these reward values, progressively improving its capability to generate high-quality evolution directions. Through the RL stage, the model not only better fits historical successful trajectories but also explores novel evolution directions, further enhancing the sustained evolutionary capability of AI systems.

\paragraph{Incremental reward model update.}
Moreover, to prevent the new policy from generating evolution directions that deviate excessively from historical successful trajectories --- which would render the reward model's feedback unreliable --- we introduce an incremental update strategy for the reward model (cf. Appendix~\ref{app:impl}). Specifically, we periodically sample recent policy outputs, execute actual verification for a subset, and when persistent discrepancies emerge between true outcomes and predictions, we augment a calibration set with these verified samples and update the reward model before resuming policy training. Through this incremental recalibration, the reward model maintains reliable feedback as the distribution shifts.

\subsection{Application to Continuous Evolution}
\label{sec:application}

\begin{figure}[t]
\centering
\includegraphics[width=\linewidth]{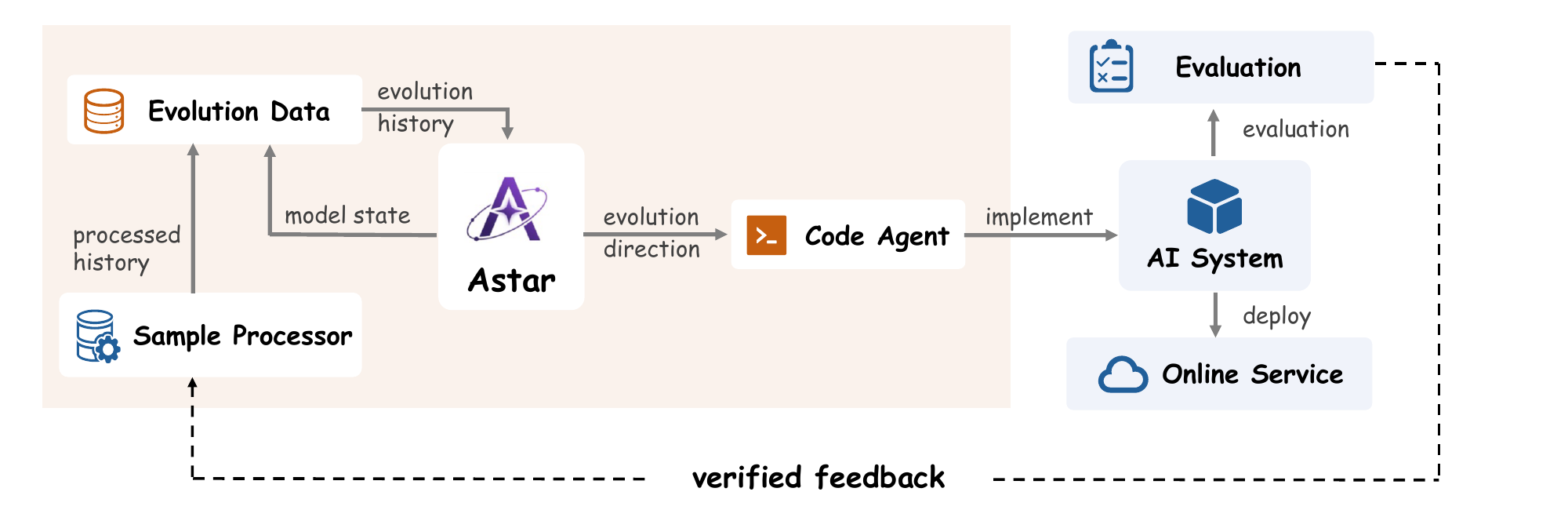}
\caption{Astar's role in continuous AI system evolution. The workflow comprises six stages: (1) Astar generates an evolution direction from the system context; (2) the Code Agent implements it as runnable code; (3) the evolution direction is trained, evaluated offline, and deployed online if successful; (4) the Sample Processor collects the results and applies noise filtering to extract structured samples; (5) verified samples feed back to continue training Astar, enabling its evaluation capability to grow as the system evolves.}
\label{fig:continuous_evolution}
\end{figure}

The training pipeline described above produces a specialized model capable of proposing effective evolution directions. However, optimizing an industrial AI system is inherently a continuous, multi-step process. Once a verified change is deployed, the system moves to a new state, and the next decision must build on it. This section describes how Astar integrates into a continuous evolution workflow that sustains performance improvements across multiple consecutive iterations.

\paragraph{The continuous evolution workflow.}
Figure~\ref{fig:continuous_evolution} illustrates how Astar participates in the continuous evolution of AI systems. The workflow operates in six stages: (1) \textbf{Direction generation}, where Astar generates an evolution direction based on the current system context, including data features and basic information about the existing codebase; (2) \textbf{Code implementation}, where a code agent (built on general-purpose LLMs such as Qwen or Claude) translates the evolution direction into runnable code modifications; (3) \textbf{Execution and deployment}, where the modified system undergoes training and offline evaluation, and successful changes are deployed to production; (4) \textbf{Sample collection and processing}, where the evolution attempt and its performance outcome are recorded, and the noise filtering pipeline (cf. Section~\ref{sec:data_denoising}) extracts structured training samples with hierarchical hints; and (5) \textbf{Feedback}, where verified samples are used to continue training Astar, thereby refining its capability to propose effective directions for future iterations.

\paragraph{Benefits of the closed loop.}
By recycling each iteration's outcomes back into training, Astar evolves alongside the target system. Without such feedback, a static Astar would fall behind as the system advances: once the system evolves beyond training states, Astar's proposals become outdated and fail to explore the new optimization frontier. This co-evolution enables sustained improvements across multiple consecutive iterations, as demonstrated by the 20-iteration deployment in Section~\ref{sec:online_exp}.

%% file: sections/5_experiments.tex
\section{Experiments}
\label{sec:results}

Through the three-stage training pipeline on the structured evolutionary corpus, we obtain \textbf{Astar}, a specialized model capable of proposing effective evolution directions by learning experience from the target system's iteration history. In this section, we conduct comprehensive experiments to evaluate Astar's performance in generating high-quality evolution directions, while also conducting several ablation studies to validate our design rationale.



\subsection{Experimental Setup}

\paragraph{Baselines.}
We compare Astar against two categories of baselines that represent current practice in AI system evolution. The first baseline consists of evolution directions proposed by senior algorithm engineers with extensive domain expertise, which represent the human expert level that current industrial systems rely upon. The second category comprises mainstream general-purpose LLMs that generate evolution directions under the same system context: GLM-5.2~\citep{GLM5-2}, Kimi-K2.6~\citep{Kimi-K2.6}, MiniMax-M2.7~\citep{MiniMax-M2.7}, DeepSeek V4 Pro~\citep{DeepSeek-V4}, Claude-4.8-Opus~\citep{Claude-4.8}, Qwen-3.7 Max~\citep{qwen3}, and GPT-5.5~\citep{gpt5}. These general-purpose LLMs are trained on broad public corpora but have no access to the target system's iteration history, which is precisely the domain experience gap that Astar addresses through training on historical evolution trajectories.

\paragraph{Evaluation metrics.}
We adopt two pass@$k$-style metrics, RM@$k$ and S@$k$, to evaluate the quality of generated evolution direction candidates. RM@$k$ is based on the offline reward model's predicted probability of whether the loss will decrease: if the prediction exceeds a threshold (e.g., 0.5 in our experiments), the candidate is considered positive; otherwise, it is considered negative. In contrast to RM@$k$ as a surrogate metric, S@$k$ uses the actual loss decrease obtained from real execution as the ground-truth but more expensive feedback for determining positive or negative candidates. Following the pass@$k$ convention~\citep{qwen3,GLM5-2}, for each system context we sample $k$ evolution direction candidates independently and count the context as successful if at least one candidate is positive, reporting the fraction of successful contexts. Experiments show that the two types of metrics exhibit high consistency (cf. Table~\ref{tab:gen_eval}).

\paragraph{Training configuration.}
We train Astar at three model scales (Astar-0.6B, Astar-4B, and Astar-8B) using Qwen3 as the base architecture. Following the three-stage training pipeline described in Section~\ref{sec:post-training}, we perform mid-training on 112B tokens, supervised fine-tuning on 81.1M tokens of positive samples, and reinforcement learning with GRPO~\citep{grpo} on 8.06M tokens. All stages share the unified input format shown in Figure~\ref{fig:data_format} with a maximum sequence length of 8192 tokens. Detailed optimization settings are provided in Appendix~\ref{app:impl}.

\subsection{Generation Quality Comparison}
Table~\ref{tab:gen_eval} presents the generation quality comparison between Astar and baselines across both offline reward model evaluation (RM@$k$) and real-execution evaluation (S@$k$). As revealed, Astar-8B achieves a single-proposal success rate (S@1) of 0.6786, substantially outperforming human experts (0.3229) and the strongest general-purpose LLM GPT-5.5 (0.3071) by a large margin. Even the smallest Astar-0.6B model reaches S@1 of 0.5435, already surpassing all baselines. This clearly demonstrates that training on the target system's iteration history enables Astar to internalize domain-specific evolution patterns that are inaccessible to general-purpose models trained only on public corpora, thereby directly addressing the domain experience gap. Notably, the high consistency between RM@$k$ and S@$k$ metrics validates the reward model as a reliable surrogate evaluator, enabling efficient large-scale candidate screening while maintaining alignment with actual performance gains.

\begin{table}[t]
\caption{Generation quality across models under the pass@$k$ convention. RM@$k$ judges effectiveness by the reward model (predicted loss decrease); S@$k$ judges it by real execution (actual loss decrease vs.\ baseline).}
\label{tab:gen_eval}
\centering
\small
\renewcommand{\arraystretch}{1.05}
\begin{tabular*}{\linewidth}{@{\extracolsep{\fill}} l cc @{\hskip 20pt} cc}
\toprule
\multirow{2}{*}{\textbf{Model}} & \multicolumn{2}{c}{\textbf{Reward Model}} & \multicolumn{2}{c}{\textbf{True Execution}} \\
\cmidrule(lr){2-3} \cmidrule(lr){4-5}
& \textbf{RM@1} & \textbf{RM@3} & \textbf{S@1} & \textbf{S@3} \\
\midrule
\modelicon{human} Human & 0.2041 & 0.4317 & 0.3229 & 0.4873 \\
\midrule
\modelicon{glm} GLM-5.2 & 0.0886 & 0.1772 & 0.1946 & 0.2353 \\
\modelicon{kimi} Kimi-K2.6 & 0.1392 & 0.2405 & 0.2138 & 0.3571 \\
\modelicon{MiniMax} MiniMax-M2.7 & 0.0633 & 0.1392 & 0.1128 & 0.1875 \\
\modelicon{deepseek} DeepSeek V4 Pro & 0.1139 & 0.1899 & 0.2182 & 0.2857 \\
\modelicon{claude} Claude-4.8-Opus & 0.1772 & 0.3357 & 0.2581 & 0.4286 \\
\modelicon{QWen} QWen-3.7 Max & 0.1646 & 0.2785 & 0.2329 & 0.3913 \\
\modelicon{ChatGPT} GPT-5.5 & 0.1899 & 0.3612 & 0.3071 & 0.4615 \\
\midrule
\textbf{Astar-0.6B} & 0.6741 & 0.8162 & 0.5435 & 0.6818 \\
\textbf{Astar-4B} & 0.7058 & 0.8616 & 0.6338 & 0.7692 \\
\textbf{Astar-8B} & \textbf{0.7183} & \textbf{0.8755} & \textbf{0.6786} & \textbf{0.7857} \\
\bottomrule
\end{tabular*}
\end{table}

\subsection{Training Stage Ablations}
To understand the contribution of each training stage, we conduct a progressive ablation study by incrementally adding mid-training, SFT, and RL to the base model. As shown in Table~\ref{tab:stage_effect}, the base Qwen3-8B model without training achieves only RM@1 of 0.0380, comparable to general-purpose LLMs. Mid-training yields the largest single improvement, boosting RM@1 to 0.6041, as it injects domain-specific evolution knowledge and the structured output conventions (cf. Figure~\ref{fig:data_format}) that enable hierarchical hint generation. SFT further lifts RM@1 to 0.6853 by aligning the output distribution toward high-quality evolution directions that demonstrably yield performance gains. Finally, RL pushes RM@1 to 0.7183, enhancing the model's exploration capability toward novel effective directions beyond the training data. This progressive improvement validates the design rationale of the three-stage training pipeline presented in Section~\ref{sec:post-training}.

\begin{table}[t]
\caption{Effect of different training stages on generation quality.}
\label{tab:stage_effect}
\centering
\small
\renewcommand{\arraystretch}{1.20}
\begin{tabular}{lcc}
\toprule
\textbf{Model} & \textbf{RM@1} & \textbf{RM@3} \\
\midrule
\textbf{Base} & 0.0380 & 0.1266 \\
\textbf{Astar-8B (mid-training)} & 0.6041 & 0.6947 \\
\textbf{Astar-8B (mid-training + SFT)} & 0.6853 & 0.8179 \\
\textbf{Astar-8B (mid-training + SFT + RL)} & \textbf{0.7183} & \textbf{0.8755} \\
\bottomrule
\end{tabular}
\end{table}

\subsection{Scalability}
To examine how generation quality scales with model size, we train Astar at three scales: 0.6B, 4B, and 8B parameters. As shown in Table~\ref{tab:gen_eval}, both RM@$k$ and S@$k$ consistently improve with model size, with Astar-8B achieving the highest S@1 of 0.6786. Notably, even the smallest Astar-0.6B already reaches S@1 of 0.5435, substantially surpassing all general-purpose LLMs and human experts. This demonstrates that domain-specific training on iteration history delivers substantial gains even at small scale~\citep{hoffmann2022training,kaplan2020scaling}, while larger models further enhance performance by retaining richer domain evolution experience. This scaling property enables flexible deployment where size can be traded against inference cost based on resource constraints.

\subsection{Data-Design Ablations}
To validate the effectiveness of key design choices in the evolutionary corpus construction pipeline (cf. Figure~\ref{fig:data_format} in Section~\ref{sec:data}), we conduct ablation studies on several critical components:

\begin{itemize}[topsep=3pt]
    \item For the input context format, as shown in Table~\ref{tab:app_ablation}(a), feeding the generation model with \textit{Basic Information} as the input context (i.e., natural-language abstractions of the codebase) achieves S@1 of 0.6786, substantially outperforming raw code repositories (0.4151). This empirically validates that abstracting code into structured natural-language descriptions reduces learning difficulty and enables better generalization at the current data scale.

    \item For the output candidate format, as shown in Table~\ref{tab:app_ablation}(b), removing \textit{Hierarchical Hints} degrades reward AUC from 0.8487 to 0.7910. This demonstrates that the hierarchical hints provide additional structural guidance that helps the model identify more effective evolution directions among a vast direction space.
\end{itemize}

\begin{table}[t]
\centering
\caption{Ablations on two data-design choices: (a) basic information vs. raw code as the input context (evaluated by real execution S@$k$); (b) hierarchical hints in the output candidate (evaluated by reward model's AUC and Accuracy).}
\label{tab:app_ablation}
\begin{minipage}[t]{0.44\linewidth}
\centering
\small
\renewcommand{\arraystretch}{1.20}
\begin{tabular}{lcc}
\toprule
\textbf{Input} & \textbf{S@1} & \textbf{S@3} \\
\midrule
\textbf{Raw code repo.} & 0.4151 & 0.5882 \\
\textbf{Basic information} & \textbf{0.6786} & \textbf{0.7857} \\
\bottomrule
\end{tabular}
\vspace{0.3em}

\centering
\small
(a) Input context format.
\label{tab:match_ablation}
\end{minipage}
\hfill
\begin{minipage}[t]{0.52\linewidth}
\centering
\small
\renewcommand{\arraystretch}{1.20}
\begin{tabular}{lcc}
\toprule
\textbf{Output} & \textbf{AUC} & \textbf{Accuracy} \\
\midrule
\textbf{w/o hierarchical hints} & 0.7910 & 0.7031 \\
\textbf{Full output} & \textbf{0.8487} & \textbf{0.7890} \\
\bottomrule
\end{tabular}
\vspace{0.3em}

\centering
\small
(b) Output candidate format.
\label{tab:rm_input_ablation}
\end{minipage}
\end{table}

\subsection{Online Experiments}
\label{sec:online_exp}

While offline metrics serve as reliable proxies, the ultimate validation of Astar lies in its deployment within a production system. To this end, we deployed Astar to continuously propose evolution directions for the online recall model of Alibaba's Lazada advertising platform. Over a two-week period, Astar guided 20 consecutive evolution iterations, achieving a cumulative 23.6\% improvement in offline Hitrate@200. Following this, we conducted an online A/B test from July 1 to July 5, 2026, with live traffic split randomly between the treatment and control groups. As Table~\ref{dataset2} reports, the treatment group driven by Astar improved on every key business metric: GMV +4.86\%, Advertising Revenue +1.82\%, Click +0.84\%, and Order +1.79\%.

These results are significant in two respects. First, the performance gains remained stable throughout the entire test window, confirming that Astar discovers genuinely effective evolution directions rather than exploiting transient statistical anomalies. Furthermore, since the recall stage fundamentally bounds downstream ranking and conversion, these improvements may compound further along the entire recommendation pipeline. Second, beyond sheer accuracy, our Astar fundamentally alters the economics of AI system evolution: it increases the throughput of verified evolution directions by over an order of magnitude while significantly reducing the labor cost from hours to minutes per direction (cf. Table~\ref{tab:velocity}), which enables rapid, continuous evolution of industrial AI systems compared to the traditional manual workflow.

\begin{table}[t]
    \centering
    \caption{Results of the online A/B experiment, with all performance gains being statistically significant at $p < 0.05$.}
    \renewcommand{\arraystretch}{1.20}
    \label{dataset2}
    \small
    \begin{tabular}{cccc}
        \toprule
            GMV  & Advertising Revenue  & Click  & Order \\
        \midrule
            +4.86\% & +1.82\% & +0.84\% & +1.79\% \\
        \bottomrule
    \end{tabular}
\end{table}

\begin{table}[t]
\caption{Evolution efficiency comparison: traditional manual workflow vs. our Astar-guided workflow.}
\label{tab:velocity}
\centering
\small
\renewcommand{\arraystretch}{1.20}
\begin{tabular*}{\linewidth}{@{\extracolsep{\fill}} l cc}
\toprule
\textbf{Metric} & \textbf{Manual Workflow} & \textbf{Astar-guided Workflow} \\
\midrule
Verified direction throughput & $\Theta(1)$--$\Theta(10)$ directions/week & $\Theta(100)$ directions/week \\
Labor cost per direction & $\Theta(1)$--$\Theta(10)$ hours/direction & $\Theta(1)$--$\Theta(10)$ minutes/direction \\
\bottomrule
\end{tabular*}
\end{table}

\subsection{Case Studies}
\label{sec:case_study}

To illustrate Astar's ability to propose high-quality, targeted evolution directions, we present two representative cases in generative retrieval scenarios (cf. Appendix~\ref{app:cases} for details).

\paragraph{Case 1: Level-specialized decoder feed-forward network.} Astar identifies that the baseline shared feed-forward network (FFN) in the decoder forces a single set of parameters to fit both coarse-grained and fine-grained codebook token distributions simultaneously, causing interference across levels. It recommends replacing the shared FFN with a mixture-of-experts architecture that assigns one dedicated expert per codebook level, with deterministic routing by level index. This structural modification remains FLOPs-neutral (each token still activates exactly one expert) while eliminating cross-level interference, demonstrating Astar's ability to capture domain-specific properties --- in this case, the level-wise semantic heterogeneity inherent to the RQ-VAE codebook structure.

\paragraph{Case 2: Spectral denoising for the Muon optimizer.} Astar recognizes that the baseline Newton-Schulz orthogonalization in the Muon optimizer normalizes all singular values to 1, thereby amplifying noise directions that were naturally suppressed by small singular values. It proposes inserting a signal/noise separation step before orthogonalization, using the Marchenko-Pastur bound from random matrix theory as a deterministic threshold: directions above the bound are orthogonalized normally, while directions within the noise distribution are attenuated. This modification preserves Muon's core benefit of equalizing signal directions while eliminating its amplification of noise, requiring no learnable parameters and adding no training burden. This case highlights Astar's capacity to understand the mathematical nature of training algorithms and propose theoretically grounded improvements.

%% file: sections/7_conclusion.tex
\section{Conclusion and Future Work}

We present \textbf{Astar}, a training-based framework that learns to propose effective evolution directions from AI systems' own iteration history. Astar addresses four key challenges through targeted designs: pairwise data expansion tackles limited direct supervision signals (C1), noise filtering removes performance-irrelevant samples (C2), hierarchical hints constrain the vast evolution direction space (C3), and a trained reward model provides fast surrogate evaluation to replace expensive real verification (C4). After three-stage training, including mid-training, SFT, and RL, Astar internalizes domain-specific evolution patterns and can propose effective directions in seconds. Deployed on Alibaba's Lazada advertising platform, Astar-8B achieves a single-proposal success rate of 0.6786, substantially outperforming human experts (0.3229) and the strongest general-purpose LLM (0.3071). Over 20 consecutive iterations, Astar improved offline Hitrate@200 by 23.6\% and yielded relative lifts of 4.86\% in GMV and 1.82\% in advertising revenue in online A/B testing. By automating the proposal of evolution directions, the most critical yet least automated stage in AI system iteration, Astar eliminates the bottleneck that previously relied on senior experts, dramatically reduces iteration cost, and enables sustainable evolution of industrial AI systems.

\paragraph{Future work.}
Looking forward, several directions can further enhance Astar's capability. First, \emph{self-improvement} through multi-round reflection and multi-agent co-judgment could reduce trial-and-error costs. Currently, Astar generates directions in a single forward pass. Multi-round self-reflection would enable iterative refinement by considering potential failure modes before committing to a direction. Furthermore, multiple specialized agents trained on different roles (e.g., innovator, critic, verifier, etc.) could collaboratively judge candidates, leading to more robust and well-rounded proposals. Such deliberation shifts computational cost from expensive real-world verification to cheaper test-time reasoning, improving success rate while maintaining rapid iteration. Second, we are extending Astar to additional scenarios beyond e-commerce advertising, including content-based recommendation, foundation model fine-tuning, and agent harnessing, to validate its generalization across diverse domains.

%% file: sections/G_data.tex
%

\section{Corpus Construction Details}
\label{app:data}

This appendix collects the detailed field schema, filtering rules, and taxonomy structure behind the evolutionary corpus of Section~\ref{sec:data}.

\subsection{Raw Experiment Record}
Each raw record is a complete snapshot of one experiment, including engineering-code snapshot, training log and metrics, and experiment metadata. Table~\ref{tab:raw_record} lists the full field structure.

\begin{table}[htbp]
\centering
\caption{Field structure of a raw experiment record.}
\label{tab:raw_record}
\small
\renewcommand{\arraystretch}{1.20}
\begin{tabular}{@{}>{\raggedright\arraybackslash}p{0.30\linewidth} >{\raggedright\arraybackslash}p{0.66\linewidth}@{}}
\toprule
\textbf{Field} & \textbf{Description} \\
\midrule
\multicolumn{2}{@{}l}{\textbf{(1) Engineering-code snapshot}} \\
\addlinespace[2pt]
Complete training-code repository & All source code packaged at submission time, including model definitions, loss functions, feature processing, training scripts, and configuration files \\
\midrule
\multicolumn{2}{@{}l}{\textbf{(2) Training log and metrics}} \\
\addlinespace[2pt]
loss sequence & Per-step loss \\
metrics & Parsed from the [Metric] lines (distinguished by MODE.TRAIN/EVAL) and the FINAL AUC output: global\_step, local\_step, loss, clk\_auc, atc\_auc, ord\_auc, qps \\
\midrule
\multicolumn{2}{@{}l}{\textbf{(3) Experiment metadata}} \\
\addlinespace[2pt]
user\_id / user\_dept & Submitter id and department \\
git\_repo / git\_branch / git\_commit & Code repository, branch, and native commit hash \\
experiment\_id & Training-platform instance id (application\_id) \\
job\_name & Human-readable task name \\
task\_type & TRAIN / EVAL / PREDICT \\
status & success / failed / killed \\
submitted\_at / started\_at / finished\_at & Submission, start, and finish timestamps (ISO8601) \\
duration\_s & Run duration (seconds) \\
runner & Entry script \\
config\_file & Configuration-file path \\
user\_define\_cmd & Complete launch command \\
model\_name & Model class name \\
docker\_image & Runtime image \\
dataset\_uri & Data-warehouse table path of the training data \\
dw\_project & Data-warehouse project the dataset belongs to \\
ckpt\_dir & Checkpoint path \\
logs\_url & Log entry link \\
\bottomrule
\end{tabular}
\end{table}

\subsection{Configuration Whitelisting Rules}
Table~\ref{tab:config_key_class} lists the complete classification rules for configuration parameters.

\begin{table}[htbp]
\centering
\caption{Configuration parameter classification rules.}
\label{tab:config_key_class}
\small
\renewcommand{\arraystretch}{1.20}
\begin{tabular}{@{}>{\raggedright\arraybackslash}p{0.24\linewidth} >{\raggedright\arraybackslash}p{0.72\linewidth}@{}}
\toprule
\textbf{Group} & \textbf{Keys (leaf key name)} \\
\midrule
\multicolumn{2}{@{}l}{\textbf{discard keys} (no training significance)} \\
\addlinespace[2pt]
Path / experiment identifier & output\_dir, platform\_model, EXPNAME, job\_name, trace\_table, trace\_partition, table\_name, partition \\
Platform resource & GPU, QUEUE, PLATFORM\_PROJECT, DW\_PROJECT, worker\_count, access\_id, access\_key, user\_id, replica, max\_failover\_times, interval\_steps, max\_to\_keep \\
Data / IO endpoint & tables, dw\_table, dw\_endpoint, tunnel\_end\_point, tunnel\_endpoint, endpoint, DW\_TABLE, CONFIG\_FILE, read\_threads\_num, save\_io\_interval \\
Command string & node\_name, engine, algo\_name, user\_params, ENV \\
Data partition path & train\_data, test\_data \\
Subtree prefix & checkpoint.model\_bank, tracer. \\
\midrule
\multicolumn{2}{@{}l}{\textbf{retain keys} (passed to the LLM for extraction)} \\
\addlinespace[2pt]
Training hyperparameter & learning\_rate, batch\_size, optimizer, weight\_decay, dropout, epoch, max\_steps, gradient\_accumulation\_steps, auc\_bucket\_num \\
Model structure / feature & model\_options, dnn\_blocks, emb\_blocks, seq\_blocks, model\_name, fg\_path \\
Unknown key & Any key not listed above (will be conservatively retained) \\
\bottomrule
\end{tabular}
\end{table}

\subsection{Syntax Normalization and Reachability Analysis}
For Python files, text-level diffs cannot reliably distinguish logical changes from superficial edits (e.g., print statements, docstrings, import reordering). We parse each file into an abstract syntax tree (AST)~\cite{allamanis2018survey,alon2019code2vec,feng2020codebert}, remove debugging statements and comments, normalize import ordering, and compare the resulting structures. Identical normalized ASTs indicate no logical change. Files that fail to parse fall back to line-level comparison after removing blank lines, comments, and print statements. We then perform reachability analysis from the script entry point. Changed functions or classes unreachable from execution paths are dead code and marked as pseudo-changes.

\subsection{Evolution-Intent Hierarchical Hints}

\begin{table}[h]
\centering
\caption{First- and second-level structure of the Evolution-Intent Hierarchical Hints.}
\label{tab:taxonomy}
\small
\renewcommand{\arraystretch}{1.20}
\begin{tabular}{@{}>{\raggedright\arraybackslash}p{0.16\linewidth} >{\raggedright\arraybackslash}p{0.80\linewidth}@{}}
\toprule
\textbf{First-level label} & \textbf{Second-level labels} \\
\midrule
Model structure & attention mechanism, attention residual, normalization, residual connection, FFN/MLP, TokenMixer, feature mixing, encoder aggregation, decoder, parameter sharing, multi-token prediction, LazyAR inference acceleration, regularization structure, embedding, user behavior sequence modeling, feature crossing, multi-task tower, two-tower retrieval, generative retrieval \\
\midrule
Loss design & Focal loss, contrastive learning, MTP auxiliary loss, multi-level token weighting, auxiliary loss, label smoothing, ranking loss, multi-task loss weighting, sample weighting \\
\midrule
Optimization strategy & Muon optimizer, optimizer replacement, second-order/preconditioned optimizer, learning-rate scheduling, gradient processing, weight decay, sparse embedding optimizer, parameter-grouped learning rate \\
\midrule
Training engineering & evaluation pipeline, evaluation metrics, experiment management, multi-round training, memory and throughput, training-side regularization, distributed training, retrieval index service, checkpoint and initialization \\
\midrule
RL strategy & policy-variant comparison, policy gradient, advantage function, reward processing, KL penalty, sampling strategy \\
\midrule
Data processing & data source switch, data sampling, feature engineering, negative sample generation, feature preprocessing and transformation, sequence sample construction, data cleaning \\
\midrule
Code refactoring & framework migration, bug fix, code simplification \\
\bottomrule
\end{tabular}
\end{table}

The predefined three-level hints classifies purified code modifications along a hierarchical structure: (Level-1) \textit{Coarse-Grained Direction} $\rightarrow$ (Level-2) \textit{Fine-Grained Modification Action} $\rightarrow$ (Level-3) \textit{Concrete Modification Plan}. Level-1 labels correspond to major technical directions, such as model structure, loss design, and optimization strategy; Level-2 labels specify particular modules or methods under each direction, such as attention mechanisms, learning-rate scheduling, and data sampling; Level-3 labels capture the concrete action of a change, such as gated-attention replacement, cosine-annealing period adjustment, and hard-sample oversampling. These hierarchical hints effectively filter out irrelevant changes while providing Astar with dense, multi-granular supervision signals during training, guiding it to progressively narrow the search space during inference. Table~\ref{tab:taxonomy} lists the first- and second-level structure.

%% file: sections/B_implementation.tex
\section{Training Details}
\label{app:impl}

This appendix provides the complete training and implementation details for Astar's three-stage training pipeline. Table~\ref{tab:train-hparams} specifies the hyperparameters across mid-training, supervised fine-tuning (SFT), reward-model training, and reinforcement learning (GRPO) stages. Algorithm~\ref{alg:rm_audit} details the periodic auditing and incremental refresh strategy for the reward model, which maintains reliable feedback as the policy distribution shifts by sampling recent rollouts, executing real environment verification, comparing predicted scores with true outcomes, and incrementally updating the reward model when persistent discrepancies emerge.

\begin{table}[htbp]
\caption{Training hyperparameters across the Mid-Training, SFT, reward-model, and RL (GRPO) stages. Symbols follow the definitions in the corresponding sections.}
\label{tab:train-hparams}
\centering
\small
\renewcommand{\arraystretch}{1.2}
\begin{tabular}{lll}
\toprule
\textbf{Stage} & \textbf{Hyperparameter} & \textbf{Value} \\
\midrule
\multirow{3}{*}{Common}
 & Base model & Qwen3-0.6B / 4B / 8B \\
 & Max sequence length & 8192 \\
 & Precision & Mixed precision \\
\midrule
\multirow{3}{*}{Mid-Training}
 & Loss scope & NTP over full sequence \\
 & Data filtering & None (full corpus) \\
 & Training tokens & $\sim$112\,B tokens \\
\midrule
\multirow{6}{*}{SFT}
 & Epochs & 5 \\
 & LR schedule & Cosine, $10^{-5} \!\rightarrow\! 10^{-6}$
 \\
 & Weight decay & $1 \times 10^{-6}$ \\
 & Gradient clipping & Enabled (max-norm 1.0) \\
 & Loss scope & Labels + evolution idea (prompt masked) \\
 & Training data &  $\sim$81.1\,M tokens \\
\midrule
\multirow{3}{*}{Reward Model}
 & Architecture & LLM backbone + mean pooling + MLP head \\
 & Max input length & 8192 \\
 & Positive ratio & $\approx 0.55$ \\
\midrule
\multirow{3}{*}{RL (GRPO)}
 & Initialization & SFT checkpoint \\
 & Group size & 16 \\
 & Training tokens & $\sim$8.06\,M tokens \\
\bottomrule
\end{tabular}
\end{table}

\begin{algorithm}[htbp]
\caption{Incremental Reward Model Update Strategy}
\label{alg:rm_audit}
\small
\begin{algorithmic}[1]
\State \textbf{Input:} SFT-initialized policy $\pi_\theta$, reward model $r_\phi$, audit interval $T_{\text{audit}}$, refresh threshold $\tau$, patience $M$
\State \textbf{Initialize:} mismatch counter $c \gets 0$, audit set $\mathcal{D}_{\text{audit}} \gets \emptyset$

\While{RL training not converged}
    \State Generate candidates with current policy $\pi_\theta$
    \State Score candidates with reward model $r_\phi$
    \State Update policy using the reward model scores

    \If{current step $\bmod T_{\text{audit}} = 0$}
        \State Sample a small batch from recent rollouts
        \State Execute the sampled candidates in the real environment
        \State Compare reward model scores with true execution outcomes

        \If{the mismatch exceeds threshold $\tau$}
            \State $c \gets c + 1$
            \State Add audited samples into $\mathcal{D}_{\text{audit}}$
        \Else
            \State $c \gets 0$
        \EndIf

        \If{$c \ge M$}
            \State Incrementally update $r_\phi$ with $\mathcal{D}_{\text{audit}}$
            \State Clear $\mathcal{D}_{\text{audit}}$
            \State Reset $c \gets 0$
        \EndIf
    \EndIf
\EndWhile
\end{algorithmic}
\end{algorithm}

%% file: sections/D_cases.tex
\section{Case Studies}
\label{app:cases}

\lstdefinestyle{pycode}{
  language=Python,
  basicstyle=\ttfamily\footnotesize,
  keywordstyle=\color{blue!70!black},
  commentstyle=\color{gray},
  stringstyle=\color{teal},
  numbers=left, numberstyle=\tiny\color{gray},
  breaklines=true, frame=single, showstringspaces=false,
  columns=fullflexible,
}

This appendix presents two representative cases that illustrate Astar's ability to propose effective evolution directions. Each case starts from the baseline implementation of a generative retrieval (GR) model, and through hierarchical hints and an evolution idea, Astar produces a targeted, directly deployable evolution direction. The two cases correspond to a model architecture modification and an optimizer algorithm modification, respectively.

\subsection{Case 1: Level-Specialized Decoder Feed-Forward Network}

The hierarchical hints of this case are as follows.

\begin{center}
\small
\renewcommand{\arraystretch}{1.20}
\begin{tabular}{@{}ll@{}}
\toprule
\textbf{Level} & \textbf{Hint} \\
\midrule
Level-1 & Model Architecture \\
Level-2 & Decoder Feed-Forward Network \\
Level-3 & Token Heterogeneity Modeling \\
\bottomrule
\end{tabular}
\end{center}

\paragraph{Target system and baseline structure.}
The system is a GR model for e-commerce advertising, built on an Encoder-Decoder structure~\cite{su2021rope,zhang2019rmsnorm,shazeer2020glu}. The encoder encodes the user behavior sequence into a user representation~\cite{kang2018sasrec,sun2019bert4rec}, and the decoder autoregressively generates, level by level, the $L$-level RQ-VAE codebook tokens of the target item~\cite{rajput2023tiger,oord2017vqvae,lee2022rqvae}. The codebook is organized by level: high-level tokens correspond to coarse-grained categories and low-level tokens to fine-grained attributes. In the baseline implementation, the SwiGLU feed-forward network (FFN) in each decoder layer is a single instance shared across all codebook levels.

\begin{lstlisting}[style=pycode, caption={Snippet A: the shared FFN, a single instance reused across all codebook levels.}, label={lst:base_ffn}]
class Qwen2DecoderCrossLayerKVTreeFix(nn.Module):
    def __init__(self, ...):
        self.self_attn  = Qwen2Attention(...)
        self.cross_attn = Qwen2CrossAttention(...)
        self.mlp = Qwen2MLP(hidden_size, intermediate_size)  # one FFN shared by all levels
    def forward(self, hidden_states, encode_emb, ...):
        ffn_in = cross_in + attn_output_s2
        return ffn_in + self.mlp(self.post_attention_layernorm(ffn_in))
\end{lstlisting}

Snippet B shows the training call flow in which this FFN sits: all levels share the same decoder stack (including the shared FFN), and a per-level InfoNCE~\cite{oord2018cpc} loss is applied.

\begin{lstlisting}[style=pycode, caption={Snippet B: GR training call flow (teacher forcing with per-level InfoNCE).}, label={lst:base_flow}]
def token_decoder_v2_vectorized(self, encode_fusion, item_seq_mask, ..., level_targets, ...):
    num_levels = self.token_level                    # number of codebook levels (L)
    # teacher forcing: concat BOS with ground-truth token embedding of each level
    decoder_input_parts = [bos_emb]
    for level in range(num_levels):
        level_token_emb = self.get_token_embedding_v2(level_targets[level], ...)
        if level != num_levels - 1:
            decoder_input_parts.append(level_token_emb)
    decoder_input = torch.cat(decoder_input_parts, dim=1)              # [B, L, D]
    # every level goes through the SAME decoder stack (shared FFN)
    level_hidden = self.decoder_transformer_tower(
        decoder_input, seq_mask, encode_fusion, item_seq_mask)        # [B, L, D]
    all_logits = torch.einsum('bld,lvd->blv', norm_pred, codebooks)   # [B, L, V]
    pos_logits = (norm_pred * norm_real).sum(dim=-1)                  # [B, L]
    level_loss = -pos_logits + torch.logsumexp(all_logits, dim=-1)    # per-level InfoNCE
\end{lstlisting}

\paragraph{The problem Astar identifies.}
After parsing the structure and the training call flow, Astar identifies a structural mismatch. The semantic distributions of the RQ-VAE codebook tokens across levels are inherently heterogeneous, whereas the shared FFN applies a single mapping $f_\theta(\cdot)$ to all levels. This forces one set of parameters to fit both coarse-grained and fine-grained distributions at once, causing interference across levels.

\paragraph{The proposed evolution direction.}
Astar recommends upgrading the shared FFN into a mixture-of-experts (MoE) network~\cite{shazeer2017moe,fedus2022switch,lepikhin2021gshard,jiang2024mixtral,dai2024deepseekmoe} with $L$ experts, using the codebook level of a token as a deterministic routing signal: a level-$\ell$ token activates only the $\ell$-th expert,
\[
h_\ell = \mathrm{FFN}_{\ell}\big(\mathrm{LN}(x_\ell)\big), \qquad \ell = 1, 2, \ldots, L,
\]
where routing is uniquely determined by the level $\ell$ and does not depend on any learnable gating network. This naturally avoids the routing loss and expert load-balancing issues common in MoE. The baseline and the modified structures are compared below.

\paragraph{Why it works.}
This modification achieves specialized modeling of heterogeneous distributions at minimal cost. Since each token still passes through exactly one expert, the forward computation and memory footprint remain strictly neutral (FLOPs-neutral), and only the parameter count grows linearly. Coarse-grained and fine-grained tokens can then be handled by dedicated experts without mutual interference. This evolution direction is not a generic suggestion. It rests on an accurate understanding of the target system's structure: it captures the domain-specific property of level-wise semantic heterogeneity in the codebook, and transforms it into a directly deployable structural modification that adds no extra training burden.

\begin{center}
\small
\renewcommand{\arraystretch}{1.20}
\begin{tabular}{@{}lcc@{}}
\toprule
 & \textbf{Baseline (Shared FFN)} & \textbf{Hierarchical FFN Experts} \\
\midrule
FFN instances & 1 & $L$ \\
Routing & none & deterministic by level (top-1) \\
Learnable gating & none & none \\
Per-token forward cost & $O(d\,d_{\text{ff}})$ & $O(d\,d_{\text{ff}})$ (unchanged) \\
Parameters & $P$ & $L \cdot P$ \\
\bottomrule
\end{tabular}
\end{center}
\subsection{Case 2: Spectral Denoising for the Muon Optimizer}

The hierarchical hints of this case are as follows.

\begin{center}
\small
\renewcommand{\arraystretch}{1.20}
\begin{tabular}{@{}ll@{}}
\toprule
\textbf{Level} & \textbf{Hint} \\
\midrule
Level-1 & Optimizer and Training \\
Level-2 & Gradient Orthogonalization \\
Level-3 & Gradient Denoising \\
\bottomrule
\end{tabular}
\end{center}

\paragraph{Target system and baseline structure.}
The same GR model is trained with a hybrid of the Muon and AdamW~\cite{kingma2015adam,loshchilov2019adamw,gupta2018shampoo,chen2023lion,liu2023sophia} optimizers: two-dimensional attention weight matrices~\cite{dao2022flashattention,ainslie2023gqa,shazeer2019mqa,ba2016layernorm} use Muon, whose gradients are orthogonalized by Newton-Schulz (NS) iteration before the update, while one-dimensional parameters and heterogeneous layers use AdamW. NS orthogonalization computes the zeroth power of the gradient matrix $G = U\Sigma V^\top$, namely $G(G^\top G)^{-1/2}\approx UV^\top$, which retains the directions $U, V$ and normalizes all singular values to 1.

\begin{lstlisting}[style=pycode, caption={Snippet A: Newton-Schulz iteration drives all singular values to 1, with no gradient denoising.}, label={lst:base_ns}]
def zeropower_via_newtonschulz5(G, steps=5):
    a, b, c = (3.4445, -4.7750, 2.0315)
    X = G.bfloat16()
    X = X / (X.norm(dim=(-2, -1), keepdim=True) + 1e-7)
    for _ in range(steps):
        A = X @ X.mT
        B = b * A + c * A @ A
        X = a * X + B @ X          # X -> U V^T  (all singular values driven to 1)
    return X
\end{lstlisting}

Snippet B shows the training call flow: the \texttt{p.grad} of the attention matrices enters NS orthogonalization directly, with no preprocessing of the gradient distribution in between.

\begin{lstlisting}[style=pycode, caption={Snippet B: the hybrid 2D-Muon plus 1D/heterogeneous-layer AdamW call flow.}, label={lst:base_step}]
@torch.no_grad()
def step(self, closure=None):
    self.adam_opt.step()                              # AdamW for non-matrix params
    for p in self.muon_params:                        # Muon for 2D attention matrices
        if p.grad is None:
            continue
        state = self.muon_state.setdefault(p, {'momentum_buffer': torch.zeros_like(p)})
        update = muon_update(
            p.grad.clone(), state['momentum_buffer'],  # p.grad enters NS as-is
            beta=muon_momentum, ns_steps=self.muon_ns_steps)
        p.add_(update.reshape(p.shape), alpha=-muon_lr)  # apply orthogonalized update
\end{lstlisting}

\paragraph{The problem Astar identifies.}
In training for recommendation, the gradient can be decomposed into a signal part and a noise part, which are naturally separable on the singular-value spectrum: signal directions correspond to larger singular values $\sigma_s$ and noise directions to smaller ones $\sigma_n$, with $\sigma_s \gg \sigma_n$ in the general case. The baseline NS orthogonalization normalizes all singular values to 1, so noise directions that were naturally suppressed by small singular values are forcibly raised to the same update weight as signal directions. This is the flip side of Muon's benefit of equalizing all directions on noisy data: it faithfully amplifies the noise.

\paragraph{The proposed evolution direction.}

Astar proposes upgrading the undifferentiated NS orthogonalization into a signal/noise-separated orthogonalization, using the upper bound on noise singular values given by the Marchenko-Pastur (MP) law from random matrix theory as a deterministic criterion. Let the aspect ratio of the gradient matrix be $\gamma = m/n$; the noise upper bound is
\[
\sigma_{\text{th}} = \sigma_{\text{noise}}\,(1+\sqrt{\gamma})^2 .
\]
Directions above $\sigma_{\text{th}}$ are judged as signal and orthogonalized normally ($\sigma \to 1$); directions within the MP distribution are judged as noise and attenuated rather than normalized. The baseline and the modified schemes are compared below.

\begin{center}
\small
\renewcommand{\arraystretch}{1.20}
\begin{tabular}{@{}lcc@{}}
\toprule
 & \textbf{Baseline (NS)} & \textbf{MP spectral denoising} \\
\midrule
Singular-value handling & all $\sigma \to 1$ & signal $\sigma \to 1$, noise attenuated \\
Signal/noise criterion & none & MP bound $\sigma_{\text{noise}}(1+\sqrt{\gamma})^2$ \\
Learnable parameters & none & none \\
Extra loss term & none & none \\
\bottomrule
\end{tabular}
\end{center}


\paragraph{Why it works.}
The MP bound is given analytically by the aspect ratio $\gamma$ of the gradient matrix alone. The criterion requires no learnable parameters, no extra loss term, and no gating or load balancing. Inserting a single spectral separation before orthogonalization preserves Muon's core benefit of equalizing signal directions, while eliminating its amplification of noise directions. As in Case 1, this evolution direction rests on an accurate understanding of the mathematical nature of the training algorithm, and it is a directly deployable modification that adds no training burden.
